\documentclass[preprint,amsmath,amssymb,aps]{revtex4-2}

\usepackage{graphicx}
\usepackage{dcolumn}
\usepackage{bm}
\usepackage[mathlines]{lineno}

\usepackage{stackrel}
\usepackage{wasysym}
\usepackage{esint}

\begin{document}
	
\preprint{APS/123-QED}
\title{Phase-Integral Formulation of \\ Dynamically Assisted Schwinger Pair
Production}
\author{Chul Min Kim}
\affiliation{Center for Relativistic Laser Science, Institute for Basic Science,
Gwangju 61005, Republic of Korea}
\affiliation{Advanced Photonics Research Institute, Gwangju Institute of Science
and Technology, Gwangju 61005, Republic of Korea}

\author{Alexander Fedotov}
\email{amfedotov@mail.ru}
\affiliation{National Research Nuclear University ``MEPhI'' (Moscow Engineering
Physics Institute), Kashirskoe sh.31, 115409 Moscow, Russia}

\author{Sang Pyo Kim}
\affiliation{Department of Physics, Kunsan National University, Kunsan 54150, Republic of Korea}

\date{Sep. 21, 2021}

\begin{abstract}
We present a phase-integral formulation of dynamically assisted Schwinger pair production of scalar charges to find the pair production density under a strong low-frequency field and a weak high-frequency field. The leading WKB action was the Schwinger formula determined by the constant field, whose corrections are determined by the Keldysh parameter for the oscillating field, $m\omega_{q}/qE_{q}$. We found a systematic expression of the leading WKB action as a power series in the Keldysh parameter, of which coefficients are given as integrals of the product of fields in the complex time domain. For the case of a strong constant field superimposed with a weak oscillating field, we provided explicit formulas and proposed a procedure for numerical evaluation. The presented phase-integral formulation should provide a clear simple method for quantitatively analyzing the leading-order features of dynamically assisted Schwinger pair production.
\end{abstract}
\maketitle

\section{Introduction}
One of the predictions of strong field Heisenberg-Euler
and Schwinger effective action \cite{Heisenberg1936Folgerungen,Schwinger1951gauge}
beyond the linear Maxwell theory is the spontaneous production of
electron and positron pairs in an electric field background. A direct
measurement of Schwinger pair production (SPP) in the strong electric
field will confirm the nature of vacuum structure in strong field
QED. The recent rapid development of ultra-intense lasers has opened
a window for the high-intensity field physics since Mourou and Strickland
invented CPA technology \cite{Strickland1985Compression}. In spite
of the rapid progress of the intense lasers, the time and cost for
building such ultra-intense lasers has forced researchers to find
an alternative to enhance the pair production rate with the given
intensity. Sch{\"u}tzhold, Gies, and Dunne introduced the so-called
dynamically assisted SPP (DA-SPP) which adopts a superposition of
a strong slowly varying field and a weak rapidly varying one to enhance
SPP \cite{Schuetzhold2008Dynamically}.

The explicit solution of the Dirac equation in a locally constant electric field (LOC) assisted by a fluctuating electric field goes beyond of the current theoretical study, and one has to
find an approximation scheme to calculate the leading and subleading
correction terms. The worldline instanton method, one of the widely
used methods, computes the action for a charge in a given electric
field in the complex plane of time \cite{Dunne2005Worldline,Dunne2006Worldline}.
The other method is the relativistic Wentzel-Kramers-Brillouin (WKB)
action from the field equation \cite{Marinov1977Electron}. Recently
Kim and Page has advanced the phase-integral formulation that finds
the WKB action through Cauchy integrals in the complex plane of time
or space for one-dimensional electric field with temporal or spatial
distribution \cite{Kim2007Improved,Kim2019Equivalence}. The advantage
of the phase-integral in the complex space is that the relativistic
action is entirely determined by the residues of the analytically
continued action provided that the electric field profile allows analytical
continuation in that complex space.

The phase-integral has been applied only to an electric field of single
profile. When it concerns about the matter of enhancing the SPP, it
is worth of extending the phase-integral to a field configuration
of a strong LOC field assisted by a weak fluctuating field. So it
is the purpose of this paper to extend and elaborate the phase-integral
method suitable for field configurations of superposed fields, in
particular, LOC field modulated by an oscillating field. To do this
we apply the phase-integral in the complex plane of time to a combined
field of LOC assisted by a fluctuating field, in which the leading
WKB action is the Schwinger formula in a constant field and the corrections
are determined by the Keldysh parameter for the oscillating field
and/or mixed Keldysh parameter for the constant field. This is the
Furry analog of the WKB action in a LOC field: the WKB action in the
LOC field is the unperturbative term and correction terms due to the
fluctuating field are computed as perturbative series. We advance
a systematic method for computing the WKB action in such a configuration
using the phase-integral and explicitly calculate the Schwinger pair
production in a LOC assisted by an oscillating field.

This paper is organized as follows. The phase-integral formulation
of Schwinger pair production (SPP) is recapitulated in Sec.~\ref{Sec:PI_SPP} and adapted for dynamically assisted Schwinger pair production
(DA-SPP) in Sec.~\ref{Sec:PI_DASPP}. In Sec.~\ref{Sec:DASPP_const_oscillating},
the DA-SPP by a strong constant field and a weak oscillating field
is analyzed to yield explicit formulas. In Sec.~\ref{Sec:DASPP_0},
a method for numerical evaluation is proposed. The
conclusion is given in Sec.~\ref{Sec:Conclusion}. The natural units
with $c=\hbar=4\pi\epsilon_{0}=1$ are used throughout, but $c$ and
$\hbar$ are restored occasionally for clarity.

\section{Phase-integral formulation of Schwinger pair production\label{Sec:PI_SPP}}
Before delving into the formulation, we characterize the SPP regime
by using the Keldysh parameter, denoted by $\gamma$. The parameter
was introduced by Keldysh to differentiate the regime of tunneling
ionization of atoms from the multi-photon ionization regime \cite{Keldysh1965Ionization,Popruzhenko2014Keldysh}.
It is roughly the ratio of the characteristic energy of the process, $mc^{2}$ for pair production, to the work done by the applied field over one (reduced) wavelength, $qEc/\omega$:

\begin{equation}
\gamma=\frac{mc^{2}}{qEc/\omega}=\frac{m\omega c}{qE},
\end{equation}
where $E$ is the peak electric field magnitude, and $\omega$ is
the field frequency. When $\gamma\lesssim1$, the field delivers an energy
comparable to the characteristic energy over its wavelength, and thus
pair production can occur as a non-perturbative field-driven process.
When $\gamma\gg1$, the delivered energy is so small that
the process may occur as a perturbative multi-photon process. Therefore,
$\gamma\lesssim1$ is assumed in our discussion of SPP.

Among the various formulations of SPP \cite{Dunne2009New}, the phase-integral
formulation proposed in Refs.~\cite{Kim2007Improved,Kim2019Equivalence}
 expresses the leading-order pair production probability
as a contour integral in the complex plane of time or space. In the present section, we recapitulate
the method for the SPP of charged scalar particles (charge $q$ and
mass $m$) under a time-dependent electromagnetic
(EM) potential $A^{\mu}(x^{\nu})=(0,0,0,A_{\parallel}(t))$.

Following DeWitt who studied the pair production in an expanding universe \cite{DeWitt1975Quantum}, we analyze the problem in Fourier space. The scalar field can be represented as a Fourier integral:
\begin{equation}
\Phi(\mathbf{x},t)=\int\frac{\mathrm{d}\mathbf{k}_{\perp}}{(2\pi)^{2}}\frac{\mathrm{d}k_{\parallel}}{2\pi}e^{i\left(\mathbf{k}_{\perp}\cdot\mathbf{x}_{\perp}+k_{\parallel}x_{\parallel}\right)}\phi_{\mathbf{k}}(t),
\end{equation}
where $\mathbf{k}_{\perp}$ ($k_{\parallel}$) denotes the momentum
perpendicular (parallel) to the EM field. When the integral is substituted into the Klein-Gordon equation minimally coupled to the EM field,
the equation of motion for each Fourier mode is obtained as 
\begin{equation}
\ddot{\phi}_{{\bf k}}(t)+\omega_{{\bf k}}^{2}(t)\phi_{{\bf k}}(t)=0,\label{sc eq}
\end{equation}
where 
\begin{equation}
\omega_{\mathbf{k}}^{2}(t)=m^{2}\Bigl[1+{\bf \kappa}_{\perp}^{2}+\Bigl(\kappa_{\parallel}-a_{\parallel}(t)\Bigr)^{2}\Bigr],\label{freq}
\end{equation}
with normalized dimensionless momenta $\boldsymbol{\kappa}_{\perp}=\mathbf{k}_{\perp}/m$
and $\kappa_{\parallel}=k_{\parallel}/m$, and the normalized vector
potential $a_{\parallel}(t)=qA_{\parallel}(t)/m$. The normalized
vector potential can be interpreted as the ratio of the field-induced energy to the relativistic energy scale $mc^2$ and as the inverse of the Keldysh parameter: $\left|a_{\parallel}(t)\right|_{\max}=1/\gamma$ for a sinusoidal field. When the
time variable in (\ref{sc eq}) is regarded as a space variable, the
equation is mathematically equivalent to the Schr\"{o}dinger equation
for above-barrier or underdense potenial scattering \cite{Pokrovskii1961problem}.
Solving the ordinary differential equation (\ref{sc eq})
for each $\mathbf{k}$, we can obtain the pair production probability.

Various methods have been developed to solve (\ref{sc eq}) \cite{Dunne2009New},
and the phase-integral method \cite{Heading2013introduction,Froeman2002Physical,Froeman2013Phase}
provides an integral representation of the leading-order pair density
per unit time \cite{Kim2007Improved,Kim2019Equivalence}: 

\begin{equation}
\frac{\mathrm{d}N_{\mathbf{k}}}{\mathrm{d}x^{3}}=\int\frac{\mathrm{d}^{3}\mathbf{k}}{(2\pi)^{3}}e^{-\mathcal{S}_{\mathbf{k}}},
\end{equation}
where $S_{\mathbf{k}}$, the relativistic Wentzel-Kramers-Brillouin
(WKB) instanton action symmetrized with respect to the inversion of
momentum, is given as a Cauchy integral in the complex time domain: 

\begin{equation}
S_{{\bf k}}=i\varointclockwise\mathrm{d}t\frac{1}{2}\Bigl(\omega_{{\bf k}}(t)+\omega_{-{\bf k}}(t)\Bigr).\label{wkb act}
\end{equation}
Here, $\omega_{\pm{\bf k}}(t)$ is made analytic by introducing branch-cuts
depending on the profile of $a_{\parallel}(t)$, and the contour encircles
the nearest zeros and branch cuts of $\omega_{{\bf \pm k}}(t)$ clockwise
in the complex $t$ domain \cite{Kim2007Improved}. Evaluating the
integral (\ref{wkb act}) leads to the leading-order pair density.

In SPP, the pair density decreases exponentially with $\mathbf{k}^{2}$
\cite{Kim2007Improved,Gelis2016Schwinger}, and thus the pair production
with small momenta is the most important.{{} In analogy
with quantum mechanics, a large transverse momentum raises the particle
energy much above the potential barrier to suppress the reflection \cite{Landau2013Quantum}.
Likewise, the mean number of pair production, given by $\left|\mathrm{reflection\:coefficient}\right|^{2}/\left|\mathrm{transmission\:coefficient}\right|^{2}$ in the quantum mechanics analogy,
is suppressed to an exponentially small number.} Thus, focusing on small $\mathbf{k}$,
we expand $S_{\mathbf{k}}$ in powers of $\mathbf{k}^{2}$. Since
$\omega_{\mathbf{k}}(t)$ in (\ref{freq}) is a function of $\kappa_{\perp}^{2}$
and $\kappa_{\parallel}$, $S_{\mathbf{k}}$ can be expressed as a
double power series in $\kappa_{\perp}^{2}$ and $\kappa_{\parallel}$
around $\boldsymbol{\kappa}=0$. In addition, as $S_{\mathbf{k}}$
is symmetric with respect to the inversion of $\mathbf{k}$ by construction,
the terms of the power series should have the form of $\kappa_{\perp}^{2p}\kappa_{\parallel}^{2q}$,
where $p$ and $q$ are non-negative integers. Therefore, $S_{\mathbf{k}}$
is expanded up to the order of $\kappa^{3}$ as

\begin{equation}
S_{\mathbf{k}}=S_{0}+S_{\perp}\cdot\mathbf{\kappa}_{\perp}^{2}+S_{\parallel}\cdot\kappa_{\parallel}^{2},\label{Sk_quad}
\end{equation}
and the pair density is given as 

\begin{equation}
\frac{\mathrm{d}N_{\mathbf{k}}}{\mathrm{d}x^{3}}=\int\frac{\mathrm{d}^{3}\mathbf{k}}{(2\pi)^{3}}e^{-\mathcal{S}_{\mathbf{k}}}=\frac{m^{3}}{8\pi^{3/2}}\frac{e^{-S_{0}}}{|S_{\perp}|\sqrt{|S_{\parallel}|}}.
\end{equation}
Consequently, once $S_{0}$, $S_{\perp}$, and $S_{\parallel}$ are
evaluated, the pair density can be obtained for small momenta. {For
null momenta, only $S_{0}$ is necessary, and the momentum integral
yielding the prefactor should be found by other methods.} In the next section, we adapt the phase-integral method for DA-SPP.

\section{Phase-integral formulation applied to dynamically assisted Schwinger
pair production\label{Sec:PI_DASPP}}

In DA-SPP, an extra weak high-frequency field called the assisting field ($a_{q}(t)$ with $\gamma_{q}\gg1$)
is superposed with a strong low-frequency field ($a_{c}(t)$ with
$\gamma_{c}\apprle1$) \cite{Schuetzhold2008Dynamically}. The field
$a_{q}(t)$ is too weak to induce SPP for itself, but its photon energy
is so high that it may enhance the SPP induced by the field
$a_{c}(t)$. In the present section, we apply the phase-integral formulation
introduced in Sec.~\ref{Sec:PI_SPP} to DA-SPP where the EM field
is given as $a_{\parallel}(t)=a_{c}(t)+a_{q}(t)$ with $\gamma_{q}\gg1$
and $\gamma_{c}\apprle1$. For convenience, we use a normalized time
$\tau=t\cdot(qE_{c}/m)$ and introduce 
$h_{c}(\tau)=a_{c}(t)$ and $h_{q}(\tau)=a_{q}(t)$. The normalization
factor $m/qE_{c}$ is regarded as the time for the pair production
by $a_{c}(t)$ \cite{Dunne2009New}. Then the frequency (\ref{freq}) is rewritten
as

\begin{equation}
\frac{\omega_{\mathbf{k}}(t)}{m}=\Bigl[1+h_{c}(\tau)^{2}+\alpha(\kappa_{\perp},\kappa_{\parallel},\tau)\Bigr]^{1/2},\label{wk_m}
\end{equation}
where 

\begin{equation}
\alpha(\kappa_{\perp},\kappa_{\parallel},\tau)=\kappa_{\perp}^{2}+\left[\kappa_{\parallel}-h_{c}(\tau)-h_{q}(\tau)\right]^{2}-h_{c}(\tau)^{2}.\label{alpha}
\end{equation}
Without the assisting field, the frequency $\omega_{\mathbf{k=0}}(t)/m$
reduces to $F_{c}(\tau)\equiv\sqrt{1+h_{c}(\tau)^{2}}$,
which is the leading term in evaluating $S_{\mathbf{k}}$.

For a small momentum, we can expand $\omega_{\mathbf{k}}(t)/m$ with $F_{c}(\tau)$ as the
unperturbed term and $\alpha(\kappa_{\perp},\kappa_{\parallel},\tau)$
as a perturbation. This is the phase-integral version of the Furry
picture that employs the dressed quantum state of a charge in a strong
background field ($a_{c}(t)$) to perturbatively calculate the
scattering amplitude involving the other weak field ($a_{q}(t)$)
\cite{Furry1951bound}. When $F_{c}(\tau)$ factored out, $\omega_{\mathbf{k}}(t)/m$ is expanded as 
\begin{equation}
\frac{\omega_{\mathbf{k}}(t)}{m}=F_{c}(\tau)\cdot\left(1+\frac{\alpha(\kappa_{\perp},\kappa_{\parallel},\tau)}{F_{c}(\tau)^{2}}\right)^{1/2}=\stackrel[n=0]{\infty}{\sum}\binom{1/2}{n}\alpha(\kappa_{\perp},\kappa_{\parallel},\tau)^{n}F_{c}(\tau)^{1-2n},\label{wk_m_exp}
\end{equation}
where $\alpha/F_{c}^{2}$ is taken as a small parameter, and $\binom{1/2}{n}$
is the binomial coefficient. By substituting (\ref{wk_m_exp}) into
(\ref{wkb act}), the instanton action $S_{{\bf k}}$ (\ref{wkb act})
is represented as
\begin{equation}
S_{{\bf k}}=\frac{iy}{2}\varointclockwise\mathrm{d}\tau\stackrel[n=0]{\infty}{\sum}\binom{1/2}{n}F_{c}(\tau)^{1-2n}\Bigl(\alpha(\kappa_{\perp},\kappa_{\parallel},\tau)^{n}+\alpha(\kappa_{\perp},-\kappa_{\parallel},\tau)^{n}\Bigr),\label{wkb exp}
\end{equation}
where $y=m^{2}/qE_{c}$, and the contour is determined solely by $F_{c}(\tau)$
in the spirit of the Furry picture. Such contour is exemplified in
Sec.~\ref{Sec:DASPP_const_oscillating}.

From (\ref{wkb exp}), the expansion forms of $S_{0}$, $S_{\perp}$, and $S_{\parallel}$
in (\ref{Sk_quad}) can be obtained. The coefficient $S_{0}$ is obtained by substituting $\kappa_{\perp}=\kappa_{\parallel}=0$
in (\ref{wkb exp}): 
\begin{equation}
S_{0}=S_{\mathbf{k}}(\kappa_{\perp}=\kappa_{\parallel}=0)=iy\varointclockwise\mathrm{d}\tau\stackrel[n=0]{\infty}{\sum}\binom{1/2}{n}F_{c}^{1-2n}\Bigl(2h_{c}h_{q}+h_{q}^{2}\Bigr)^{n},\label{wkb 0-act}
\end{equation}
in which the $n=0$ term equals the worldline instanton action under
the strong field alone \cite{Kim2019Equivalence}. The coefficients
$S_{\perp}$ and $S_{\parallel}$ are the coefficients of the $\kappa_{\perp}^{2}$
term in $S_{\mathbf{k}}(\kappa_{\parallel}=0)$ and the $\kappa_{\parallel}^{2}$
term in $S_{\mathbf{k}}(\kappa_{\perp}=0)$, respectively: 
\begin{equation}
S_{\mathbf{k}}(\kappa_{\parallel}=0)=iy\varointclockwise\mathrm{d}\tau\stackrel[n=0]{\infty}{\sum}\binom{1/2}{n}F_{c}^{1-2n}\Bigl(\kappa_{\perp}^{2}+2h_{c}h_{q}+h_{q}^{2}\Bigr)^{n}\label{wkb perp}
\end{equation}
and 
\begin{equation}
S_{\mathbf{k}}(\kappa_{\perp}=0)=i\frac{y}{2}\varointclockwise\mathrm{d}\tau\stackrel[n=0]{\infty}{\sum}\binom{1/2}{n}F_{c}^{1-2n}\left\{ \left[\kappa_{\parallel}^{2}+2k_{\parallel}(h_{c}+h_{q})+2h_{c}h_{q}+h_{q}^{2}\right]^{n}+\left[\kappa_{\parallel}\rightarrow-\kappa_{\parallel}\right]^{n}\right\} .\label{wkb para}
\end{equation}
Expanding $\left(\cdots\right)^{n}$ in (\ref{wkb 0-act}), (\ref{wkb perp}),
and (\ref{wkb para}), we find that $S_{0}$, $S_{\perp}$, and
$S_{\parallel}$ are given as linear combinations of the contour integrals

\begin{equation}
S_{pqr}\equiv\varointclockwise\mathrm{d}\tau h_{c}(\tau)^{p}\left[\gamma_{q}h_{q}(\tau)\right]^{q}F_{c}(\tau)^{r},\label{Spqr_general}
\end{equation}
where $p$, $q$, and $r$ are non-negative integers. Although $S_{pqr}$
has $\gamma_{q}$ in its definition, it would hardly depend on $\gamma_{q}$
as exemplified with a sinusoidal field in Sec.~\ref{Sec:DASPP_const_oscillating}.
When the resulting double summation is rearranged (See App.~\ref{appendixA}.), $S_{0}$
is given in a simple form:

\begin{equation}
S_{0}=iy\left(\stackrel[\mu\:\mathrm{odd}]{\infty}{\sum}\,\stackrel[\nu\:\mathrm{odd}]{\mu}{\sum}+\stackrel[\mu\:\mathrm{even}]{\infty}{\sum}\,\stackrel[\nu\:\mathrm{even}]{\mu}{\sum}\right)\left(\frac{1}{\gamma_{q}}\right)^{\mu}\binom{1/2}{(\mu+\nu)/2}\binom{(\mu+\nu)/2}{(\mu-\nu)/2}2^{\nu}S_{\nu\mu\left(1-\mu-\nu\right)},\label{eq:app_S0-1}
\end{equation}
where $\mu$ and $\nu$ are non-negative integers. As $S_{0}$ is
expressed as a power series in $\gamma_{q}^{-1}$, we can quantitatively
identify the perturbative contribution of the assisting field $a_{q}(t)$
order by order. The coefficients $S_{\perp}$ and $S_{\parallel}$
are given in similar forms in App.~\ref{appendixA}. 

Consequently, the pair density calculation in DA-SPP has been reduced to
evaluating the contour integrals (\ref{Spqr_general}). Furthermore,
the contribution of the assisting field can now be identified order
by order. In the next section, we present a concrete example of the
formulation, which yields explicit analytic formulas.

\section{Dynamically assisted Schwinger pair production by a strong constant field and a weak oscillating field\label{Sec:DASPP_const_oscillating}}

\begin{figure}[b]
\includegraphics[width=0.6\textwidth]{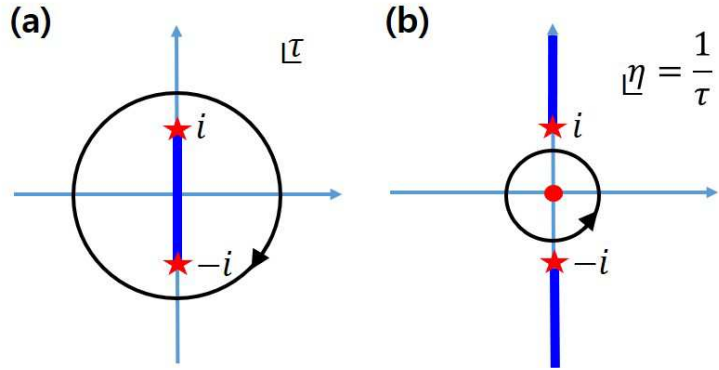}\caption{\label{fig_contours} Integration contours (a) in the complex $\tau$
space and (b) in the complex $\eta$ space, where $\eta=1/\tau$.
In (a), the red stars denote the zeros of $F_{c}(\tau)=\left(1+\tau^{2}\right)^{1/2}$,
and, in (b), the zeros of $F_{c}(1/\eta)=\left[1+\left(1/\eta\right)^{2}\right]^{1/2}$.
The thick blue lines denote the branch cuts. In (b), $F_{c}(1/\eta)$
has a singularity at the origin, denoted by a red bullet. This singularity
corresponds to the singularities at infinity in the $\tau$ space.
The contour direction is inverted by the transformation $\tau \rightarrow 1/\tau$.}
\end{figure}

In this section, we consider the DA-SPP by a strong constant field and a weak oscillating one to obtain explicit analytic expressions. In such case, the normalized vector potentials are given as

\begin{equation}
h_{c}(\tau)=\tau,\:h_{q}(\tau)=\frac{1}{\gamma_{q}}\sin(x\tau)\label{eq:const_sinu}
\end{equation}
where $x=m\omega_{q}/qE_{c}$. The contour integral $S_{pqr}$ is
given as 

\begin{equation}
S_{pqr}(x)=\varointclockwise\mathrm{d}\tau\tau^{p}\left[\sin(x\tau)\right]^{q}(1+\tau^{2})^{r/2}.
\end{equation}
The integral can be evaluated with the contour determined by the strong
constant field alone, as shown in Fig.~\ref{fig_contours}(a). The
contour encircles the zeros and the branch cut of $F_{c}(\tau)=\left(1+\tau^{2}\right)^{1/2}$. As $\tau$ and $\tau\sin\left(x\tau\right)$ are entire functions, the branch cut is not changed further. The details of evaluation is
given in App.~\ref{AppendixB}, in which $S_{pqr}(x)$ is expressed
in terms of the generalized hypergeometric functions. When the calculated
$S_{pqr}(x)$ in App.~\ref{AppendixB} are substituted in the expressions of $S_{0}$, $S_{\perp}$, and $S_{\parallel}$ in App.~\ref{appendixA}, these coefficients are obtained as a power series in $\gamma_{q}^{-1}$ involving the modified Bessel functions of the first kind, as shown in Tab.~\ref{tab:S_series}.

\begin{table}
\begin{tabular}{c|c|c|c|c}
\hline 
$n$ & $\begin{array}{c}
a_{n}(x)\\
\mathrm{in}\:S_{0}=\pi y\sum_{n}a_{n}(x)\gamma_{q}^{-n}
\end{array}$ & $a_{n}(x\rightarrow0)$ & $\begin{array}{c}
b_{n}(x)\\
\mathrm{in}\:S_{\perp}=\pi y\sum_{n}b_{n}(x)\gamma_{q}^{-n}
\end{array}$ & $\begin{array}{c}
c_{n}(x)\\
\mathrm{in}\:S_{\parallel}=\pi y\sum_{n}c_{n}(x)\gamma_{q}^{-n}
\end{array}$\tabularnewline
\hline 
\hline 
0 & $1$ & 1 & $1$ & $0$\tabularnewline
\hline 
1 & $-2I_{1}(x)$ & $-x$ & $-xI_{0}(x)$ & $x^{2}I_{1}(x)$\tabularnewline
\hline 
2 & $xI_{1}(2x)$ & $x^{2}$ & $x^{2}I_{0}(2x)$ & $-2x^{3}I_{1}(2x)$\tabularnewline
\hline 
3 & $x^{2}\left[I_{1}(x)-3I_{1}(3x)\right]/4$ & $-x^{3}$ & $x^{3}\left[I_{0}(x)-9I_{0}(3x)\right]/8$ & $x^{4}\left[-I_{1}(x)+27I_{1}(3x)\right]/8$\tabularnewline
\hline 
4 & $x^{3}\left[2I_{1}(4x)-I_{1}(2x)\right]/3$ & $x^{4}$ & $\begin{array}{c}
x^{4}\left[-I_{2}(2x)+4I_{4}(4x)\right]/3\\
+x^{3}\left[-I_{3}(2x)+8I_{3}(4x)\right]/3\\
+x^{2}\left[-2I_{2}(2x)+2I_{2}(4x)\right]/3
\end{array}$ & $2x^{5}\left[I_{1}(2x)-8I_{1}(4x)\right]/3$\tabularnewline
\hline 
\end{tabular}

\caption{\label{tab:S_series}Coefficients $S_{0}$, $S_{\perp}$, and $S_{\parallel}$
expressed in powers of $\gamma_{q}^{-1}$ up to the fourth order for
the DA-SPP by a strong constant field and a weak oscillating field.
The function $I_{n}(x)$ is the $n$th-order modified Bessel function
of the first kind, which asymptotically approaches $(x/2)^{n}/n!$ as
$x$ goes to zero \cite{Olver2010NIST}. $x=m\omega_{q}/qE_{c}$
and $y=m^{2}/qE_{c}$. }
\end{table}

The basic features of DA-SPP can be found from the expansion of $S_{0}$
in Tab.~\ref{tab:S_series}. First, the 0th-order term $\pi y$ is
the well-known exponent of the pair production formula for a constant
field, i.e., $\pi m^{2}/qE_{c}$ \cite{Heisenberg1936Folgerungen,Heisenberg2006Consequences,Schwinger1951gauge}.
Second, the sign of the first-order term $a_{1}(x)$ is negative,
implying that pair production is enhanced by the weak oscillating
field, as shown in \cite{Schuetzhold2008Dynamically}. Third, when
the frequency of the oscillating field goes to zero, i.e., $x\rightarrow0$
while $x/\gamma_{q}=E_{q}/E_{c}=\mathrm{constant}$, the expansion
reduces to $\pi m^{2}/q(E_{c}+E_{q})$, as indicated by the asymptotic
forms of $a_{n}(x)$ in Tab.~\ref{tab:S_series}. This formula is
just the lowest-order pair production formula with $E_{c}$ replaced
by $E_{c}+E_{q}$, as it should be for our formulation to be consistent.
Without the oscillating field ($h_{q}=0$), all the formulae in this
section reduce to those in \cite{Kim2019Equivalence}, in which $\delta=qE_{0}/\pi m^{2}=1/\pi y$ is used instead of $y$.

Also, the WKB instanton action obtained above is
consistent with that from the 1D Dirac equation, albeit the differences
between the governing equations. The second-order Dirac equation with
an electric field \cite{Itzykson2006Quantum} is reduced in 1D to

\begin{equation}
\left\{ \left[\left(i\partial-qA(t)\right)^{2}-m^{2}\right]I_{2\times2}-ieE(t)\sigma_{x}\right\} \left(\begin{array}{c}
\psi_{1}(t,x)\\
\psi_{2}(t,x)
\end{array}\right)=0,
\end{equation}
where $I_{2\times2}$ is the $2\times2$ identity matrix, $\sigma_{x}$
the first Pauli matrix, and $E(t)=-\dot{A}(t)$. The equation is further simplified to be diagonal by introducing spin-polarized states $\psi_{\pm}\equiv\left(\psi_{1}\pm\psi_{2}\right)/\sqrt{2}$:

\begin{equation}
\left[\left(i\partial-qA(t)\right)^{2}-m^{2}\mp ieE(t)\right]\psi_{\pm}(t,x)=0,
\end{equation}
which differs from the 3D Klein-Gordon equation by $\mp ieE(t)$.
The corresponding WKB instanton action was obtained in \cite{Linder2015Pulse}:

\begin{equation}
S_{\mathbf{k}}^{(1+1)}=\pi y\left[1-\frac{2}{\gamma_{q}}\cos\left(\kappa_{\parallel}x\right)I_{1}(x)\right],
\end{equation}
written in our notation. For small $\kappa_{\parallel}x$, $S_{\mathbf{k}}^{(1+1)}$
reduces to $S_{0}+S_{\parallel}\cdot\kappa_{\parallel}^{2}$ up to
the first order of $\gamma_{q}^{-1}$, which can be verified by using
$a_{n}(x)$ and $c_{n}(x)$ in Tab.~\ref{tab:S_series}. Since transverse
dimensions are missing in this case, $S_{\perp}=0$.
Such consistency demonstrates the generic nature of the leading-order
behavior in DA-SPP.

With the formulae of $S_{0}$, $S_{\perp}$, and $S_{\parallel}$
given in terms of $I_{n}(x)$ in Tab.~\ref{tab:S_series}, we can
estimate the pair density over a wide parameter space to understand
the behavior of DA-SPP.

\section{Method for numerical evaluation\label{Sec:DASPP_0}}

The explicit formulas obtained in the previous section would allow us to investigate the behavior of DA-SPP for a range of parameters. However, a direct evaluation of the formulas would be plagued by the divergent behavior of $I_n (x)$ in Tab.~\ref{tab:S_series}. In this section, we propose a procedure for mitigating this problem by using the Pad\'{e} approximation method, taking the evaluation of $S_0$ as a specific example. 
 
The instanton action $S_{0}$ is given as a power series
in $\gamma_{q}^{-1}$ with its coefficients depending on $x$ (Tab.~\ref{tab:S_series}):

\begin{equation}
S_{0}=\pi y\sum_{n=0}a_{n}(x)\gamma_{q}^{-n}=\pi y\cdot\hat{S}_{0}(\gamma_{q}^{-1}|x).\label{S0_exp}
\end{equation}
Note that $\hat{S}_{0}(\gamma_{q}^{-1}|x)$ approaches 1 as the assisting field diminishes ($\gamma_{q}^{-1}\rightarrow0$), indicating that $\hat{S}_{0}(\gamma_{q}^{-1}|x)$
solely represents the assisting effect. The pair density and the enhancement
due to the assisting field are given as
\begin{equation}
e^{-\pi y\hat{S}_{0}(\gamma_{q}^{-1}|x)}\label{pair_density}
\end{equation}
and
\begin{equation}
\frac{e^{-S_{0}(\gamma_{q}^{-1}\neq0)}}{e^{-S_{0}(\gamma_{q}^{-1}=0)}}=e^{\pi y\left(1-\hat{S}_{0}(\gamma_{q}^{-1}|x)\right)},\label{enhance_factor}
\end{equation}
respectively. It seems straightforward to evaluate $\hat{S}_{0}(\gamma_{q}^{-1}|x)$
by using the power series, but it is hardly so in practice. First,
it becomes formidable to obtain $a_{n}(x)$ as $n$ increases, and
thus only a limited number of terms are available in practice. Second,
the series $\sum_{n=0}a_{n}(x)\gamma_{q}^{-n}$
may diverge for large $x$ because $I_{n}(x)$ in $a_{n}(x)$ has
an asymptotic form of $I_{n}(x)\sim e^{x}/\sqrt{2\pi x}$ for $x\gg1$
\cite{Olver2010NIST}. To overcome or mitigate these difficulties, we use the Pad\'e
approximation to evaluate $\hat{S}_{0}(\gamma_{q}^{-1}|x)$.

In (\ref{S0_exp}), $\sum_{n=0}^{N}a_{n}(x)\gamma_{q}^{-n}$
may be considered as a polynomial approximation to $\hat{S}_{0}(\gamma_{q}^{-1}|x)$,
and then other forms of approximation can be used instead. In the Pad\'e
approximation method, rational functions, called Pad\'e approximants (PAs),
are used to approximate a function, and they frequently result in
finite values even when the polynomial approximations produce divergent
results \cite{Bender2013Advanced}. Furthermore, given an $N$th-order
polynomial, the PA of the type $P_{a}^{b}$ can be uniquely determined,
where $a$ ($b$) is the order of the PA's denominator (numerator),
and $N=a+b$. We employ the diagonal PA $P_{3}^{3}(\gamma_{q}^{-1}|x)$
obtained from $\sum_{n=0}^{6}a_{n}(x)\gamma_{q}^{-n}$, as shown in
App.~\ref{Appendix_Pade}:

\begin{equation}
P_{3}^{3}(\gamma_{q}^{-1}|x)=\frac{p_{0}(x)+p_{1}(x)\gamma_{q}^{-1}+p_{2}(x)\gamma_{q}^{-2}+p_{3}(x)\gamma_{q}^{-3}}{1+q_{1}(x)\gamma_{q}^{-1}+q_{2}(x)\gamma_{q}^{-2}+q_{3}(x)\gamma_{q}^{-3}},
\end{equation}
where $p_{i}(x)$'s and $q_{j}(x)$'s are calculated from $a_{n}(x)$'s.
For each value of $x$, we have a PA.

\section{Conclusion\label{Sec:Conclusion}}
We extended the phase-integral formalism developed for Schwinger pair production \cite{Kim2019Equivalence} for dynamically assisted Schwinger pair production. By combining the formalism with the Furry picture, we could systematically express the contribution of the assisting field order by order. Furthermore, we proposed a method for numerically evaluating the resulting expressions that can diverge when evaluated naively. The presented formulation should provide a clear and straightforward way for analyzing the leading-order behavior of dynamically assisted Schwinger pair production.

\begin{acknowledgments}
This work was supported by the Institute for Basic Science (IBS) under
IBS-R012-D1. The authors were benefited from the discussions during
the 28th Annual Laser Physics International Workshop (LPHYS'19) and
the 3rd Conference on Extremely High Intensity Laser Physics (ExHILP
2019). AF would like to thank the warm hospitality at CoReLS, where
this work was initiated, and he was supported by
the MEPhI Academic Excellence Project (Contract No.02.a03.21.0005) and the Russian Foundation for Basic Research (grant no. 19-02-00643).
\end{acknowledgments}

\appendix
\section{Coefficients $S_{0}$, $S_{\perp}$, and $S_{\parallel}$ as power series in $\gamma_{q}^{-1}$\label{appendixA}}

The formulas of $S_{0}$, $S_{\perp}$, and $S_{\parallel}$ can be
obtained as series by expanding $\alpha(\kappa_{\perp},\kappa_{\parallel},\tau)^{n}+\alpha(\kappa_{\perp},-\kappa_{\parallel},\tau)^{n}$
in (\ref{wkb exp}). For $S_{0}$, it is $\Bigl(2h_{c}h_{q}+h_{q}^{2}\Bigr)^{n}$
in (\ref{wkb 0-act}) to be expanded. When $\Bigl(2h_{c}h_{q}+h_{q}^{2}\Bigr)^{n}$ is expanded by the binomial expansion formula, $S_{0}$ is given as

\begin{equation}
S_{0}=iy\stackrel[n=0]{\infty}{\sum}\stackrel[m=0]{n}{\sum}\binom{1/2}{n}\binom{n}{m}2^{n-m}\varointclockwise\mathrm{d}\tau h_{c}^{n-m}h_{q}^{n+m}F_{c}^{1-2n},
\end{equation}
where $\binom{a}{b}$ denotes the binomial coefficient. This expression
shows that $S_{0}$ is a linear combination of the elemental integrals
$S_{pqr}$'s, defined as

\[
S_{pqr}\equiv\varointclockwise\mathrm{d}\tau h_{c}(\tau)^{p}\left[\gamma_{q}h_{q}(\tau)\right]^{q}F_{c}(\tau)^{r}.
\]
Consequently, once $S_{pqr}$'s are evaluated, so is $S_{0}$. After
rewriting $S_{0}$ in terms of $S_{pqr}$, we can represent $S_{0}$
as a power series in $\gamma_{q}^{-1}$ by rearranging the double
summation with new indices $\mu=n+m$ and $\nu=n-m$:

\begin{equation}
S_{0}=iy\left(\stackrel[\mu\:\mathrm{odd}]{\infty}{\sum}\,\stackrel[\nu\:\mathrm{odd}]{\mu}{\sum}+\stackrel[\mu\:\mathrm{even}]{\infty}{\sum}\,\stackrel[\nu\:\mathrm{even}]{\mu}{\sum}\right)\left(\frac{1}{\gamma_{q}}\right)^{\mu}\binom{1/2}{(\mu+\nu)/2}\binom{(\mu+\nu)/2}{(\mu-\nu)/2}2^{\nu}S_{\nu\mu\left(1-\mu-\nu\right)},\label{eq:app_S0}
\end{equation}
in which $\mu$ and $\nu$ are non-negative integers. This form of
$S_{0}$ should be convenient in studying the effect of the weak high-frequency
field $h_{q}(\tau)$. 

By the same token, $S_{\perp}$ and $S_{\parallel}$ can also be expressed
as power series in $\gamma_{q}^{-1}$:

\begin{equation}
S_{\perp}=iy\left(\stackrel[\mu\:\mathrm{odd}]{\infty}{\sum}\,\stackrel[\nu\:\mathrm{odd}]{\mu}{\sum}+\stackrel[\mu\:\mathrm{even}]{\infty}{\sum}\,\stackrel[\nu\:\mathrm{even}]{\mu}{\sum}\right)\left(\frac{1}{\gamma_{q}}\right)^{\mu}\binom{1/2}{(\mu+\nu)/2+1}\binom{(\mu+\nu)/2}{(\mu-\nu)/2}2^{\nu}\left(\frac{\mu+\nu}{2}+1\right)S_{\nu\mu\left(-1-\mu-\nu\right)}\label{eq:app_So}
\end{equation}
and

\begin{equation}
S_{\parallel}=S_{\perp}+S_{\parallel,0}+S_{\parallel,1}+S_{\parallel,2},\label{eq:app_Sp_sum}
\end{equation}
where

\begin{equation}
\begin{array}{c}
S_{\parallel,0}=iy\left(\stackrel[\mu\:\mathrm{odd}]{\infty}{\sum}\,\stackrel[\nu\:\mathrm{odd}]{\mu}{\sum}+\stackrel[\mu\:\mathrm{even}]{\infty}{\sum}\,\stackrel[\nu\:\mathrm{even}]{\mu}{\sum}\right)\left(\frac{1}{\gamma_{q}}\right)^{\mu}\binom{1/2}{(\mu+\nu)/2+2}\binom{(\mu+\nu)/2}{(\mu-\nu)/2}2^{\nu+1}\\
\times\left(\frac{\mu+\nu}{2}+1\right)\left(\frac{\mu+\nu}{2}+2\right)S_{(\nu+2)\mu\left(-3-\mu-\nu\right)},
\end{array}\label{eq:app_Sp_0}
\end{equation}

\begin{equation}
\begin{array}{c}
S_{\parallel,1}=iy\left(\stackrel[\mu\geq2,\mathrm{even}]{\infty}{\sum}\,\stackrel[\nu\:\mathrm{odd}]{\mu-1}{\sum}+\stackrel[\mu\:\mathrm{odd}]{\infty}{\sum}\,\stackrel[\nu\:\mathrm{even}]{\mu-1}{\sum}\right)\left(\frac{1}{\gamma_{q}}\right)^{\mu}\binom{1/2}{(\mu+\nu-1)/2+2}\binom{(\mu+\nu-1)/2}{(\mu-\nu-1)/2}2^{\nu+2}\\
\times\left(\frac{\mu+\nu-1}{2}+1\right)\left(\frac{\mu+\nu-1}{2}+2\right)S_{(\nu+1)\mu\left(-2-\mu-\nu\right)},
\end{array}\label{eq:app_Sp_1}
\end{equation}
and

\begin{equation}
\begin{array}{c}
S_{\parallel,2}=iy\left(\stackrel[\mu\geq3,\mathrm{odd}]{\infty}{\sum}\,\stackrel[\nu\:\mathrm{odd}]{\mu-2}{\sum}+\stackrel[\mu\geq2,\mathrm{even}]{\infty}{\sum}\,\stackrel[\nu\:\mathrm{even}]{\mu-2}{\sum}\right)\left(\frac{1}{\gamma_{q}}\right)^{\mu}\binom{1/2}{(\mu+\nu-2)/2+2}\binom{(\mu+\nu-2)/2}{(\mu-\nu-2)/2}2^{\nu+1}\\
\times\left(\frac{\mu+\nu-2}{2}+1\right)\left(\frac{\mu+\nu-2}{2}+2\right)S_{\nu\mu\left(-1-\mu-\nu\right)}.
\end{array}\label{eq:app_Sp_2}
\end{equation}

\section{Contour integral $S_{pqr}$ for a strong constant field superposed with a weak oscillating field\label{AppendixB}}

When a strong constant field is superposed with a weak oscillating field,
the contour integrals $S_{pqr}$ can be evaluated to have explicit forms. Substituting (\ref{eq:const_sinu}) in (\ref{Spqr_general}), $S_{pqr}$ is given as

\begin{equation}
S_{pqr}(x)=\varointclockwise\mathrm{d}\tau\tau^{p}\left[\sin(x\tau)\right]^{q}\left(1+\tau^{2}\right)^{r/2},\label{Spqr}
\end{equation}
of which integration contour is shown in Fig.~\ref{fig_contours}(a).
Expanding $\left[\sin(x\tau)\right]^{q}$ by using the identity $\sin^{q}\theta=\left[\left(e^{i\theta}-e^{-i\theta}\right)/2i\right]^{q}=(2i)^{-q}\stackrel[n=0]{q}{\sum}(-1)^{n}\binom{q}{n}\left[\cos\left(\theta(q-2n)\right)+i\sin\left(\theta(q-2n)\right)\right]$,
$S_{pqr}(x)$ is expressed as a linear combination of $S_{p1r}(x)$
and $C_{p1r}(x)$, where $C_{p1r}(x)\equiv\varointclockwise\mathrm{d}\tau\tau^{p}\cos(x\tau)\left(1+\tau^{2}\right)^{r/2}$:

\begin{equation}
S_{pqr}(x)=(2i)^{-q}\stackrel[n=0]{q}{\sum}(-1)^{n}\binom{q}{n}\left[C_{p1r}((q-2n)x)+iS_{p1r}((q-2n)x)\right].\label{eq:Spqr_exp}
\end{equation}
Consequently, evaluating $S_{p1r}(x)$ and $C_{p1r}(x)$ is sufficient
to obtain $S_{pqr}(x)$.

To evaluate $S_{p1r}(x)$, we make a conformal transformation $\eta=1/\tau$:

\begin{equation}
S_{p1r}(x)=-\varointctrclockwise\mathrm{d\eta}\eta^{-p-2}\sin\left(\frac{x}{\eta}\right)\left[1+\frac{1}{\eta^{2}}\right]^{r/2}=-\varointctrclockwise\mathrm{d\eta}\eta^{-p-2-r}\sin\left(\frac{x}{\eta}\right)\left(1+\eta^{2}\right)^{r/2},
\end{equation}
of which integration contour is shown in Fig.~\ref{fig_contours}(b).
Expanding $\sin(x/\eta)$ and $\left(1+\eta^{2}\right)^{r/2}$ around
$\eta=0$ ($|\eta|<1$ along the contour),

\begin{equation}
\sin\left(\frac{x}{\eta}\right)=\stackrel[n=0]{\infty}{\sum}\left(\frac{x}{\eta}\right)^{2n+1}\frac{(-1)^{n}}{(2n+1)!}\:\mathrm{and}\:\left(1+\eta^{2}\right)^{r/2}=\stackrel[n=0]{\infty}{\sum}\binom{r/2}{n}\eta^{2n},
\end{equation}
we can find the residues to complete the integration. Then, the
explicit formula of $S_{p1r}(x)$ is obtained as 

\begin{equation}
S_{p1r}(x)=-2\pi i\stackrel[n=0]{\infty}{\sum}\binom{r/2}{(p+r)/2+n+1}\frac{(-1)^{n}x^{2n+1}}{(2n+1)!},\label{eq:Sp1r}
\end{equation}
which can be written in terms of the generalized hypergeometric function $\phantom{}_{1}F_{2}(a_1;b_1,b_2;z)$:

\begin{equation}
S_{p1r}(x)=-2\pi i\binom{r/2}{(p+r+2)/2}x\cdot\phantom{}_{1}F_{2}\left(\frac{p+2}{2};\frac{3}{2},\frac{p+r+4}{2};\frac{x^{2}}{4}\right). \label{eq:S_p1r_F}
\end{equation}
By the same token, $C_{p1r}(x)$ is obtained as

\begin{equation}
C_{p1r}(x)=-2\pi i\stackrel[n=0]{\infty}{\sum}\binom{r/2}{(p+r+1)/2+n}\frac{(-1)^{n}x^{2n}}{(2n)!}\label{eq:Cp1r}
\end{equation}
or 

\begin{equation}
C_{p1r}(x)=-2\pi i\binom{r/2}{(p+r+1)/2}\phantom{}_{1}F_{2}\left(\frac{p+1}{2};\frac{1}{2},\frac{p+r+3}{2};\frac{x^{2}}{4}\right). \label{eq:C_p1r_F}
\end{equation}
Since $S_{p1r}(x)$ and $C_{p1r}(x)$ are given explicitly, so are $S_{pqr}(x)$. This evaluation technique can also be used for the case without the oscillating field, i.e., $q=0$ in (\ref{Spqr}):
\begin{equation}
	S_{p0r}(x)=-2\pi i\binom{r/2}{(p+r+1)/2}. \label{eq:S_p0r_exp}
\end{equation}

\section{Pad\'e approximant for $S_{0}$\label{Appendix_Pade}}
To calculate the Pad\'e approximant (PA) $P_{3}^{3}$, we need 7
($=3+3+1)$ terms in the polynomial $\sum_{n=0}^{M}a_{n}(x)\gamma_{q}^{-n}$.
The first 5 terms, $a_{0}(x)$ through $a_{4}(x)$, are given in Tab.~\ref{tab:S_series},
and $a_{5}(x)=x^{4}\left[-2I_{1}(x)+81I_{1}(3x)-125I_{1}(5x)\right]/192$,
and $a_{6}(x)=x^{5}\left[5I_{1}(2x)-64I_{1}(4x)+81I_{1}(6x)\right]/120$,
which are obtained from (\ref{eq:app_S0}), (\ref{eq:Spqr_exp}),
(\ref{eq:Sp1r}), and (\ref{eq:Cp1r}). The approximant $P_{3}^{3}$
is given as 

\begin{equation}
P_{3}^{3}(\gamma_{q}^{-1}|x)=\frac{p_{0}(x)+p_{1}(x)\gamma_{q}^{-1}+p_{2}(x)\gamma_{q}^{-2}+p_{3}(x)\gamma_{q}^{-3}}{1+q_{1}(x)\gamma_{q}^{-1}+q_{2}(x)\gamma_{q}^{-2}+q_{3}(x)\gamma_{q}^{-3}},
\end{equation}
where $p_{i}(x)$'s and $q_{j}(x)$'s are calculated from $a_{n}(x)$'s
according to the recurrence relations (8.3.2)-(8.3.4) in \cite{Bender2013Advanced}
or (5.12.1)-(5.12.6) in \cite{Press2007Numerical}:
{
\begin{equation}
p_{0}(x)=a_{0}(x)=1,\:q_{0}(x)\equiv1,\quad\label{eq:rec1} \\ 
\end{equation}
}
{
\begin{equation}
\sum_{m=1}^{N}q_{m}(x)a_{N-m+k}(x)=-a_{N+k}(x), \quad (k=1,\dots,N),\label{eq:rec2}
\end{equation}
}
{
\begin{equation}
\sum_{m=0}^{k}q_{m}(x)a_{k-m}(x)=p_{k}(x),\quad (k=1,\dots,N) \label{eq:rec3}
\end{equation}
for the $N$th-order diagonal PA $P_{N}^{N}(\gamma_{q}^{-1}|x)$.
Note that we have a PA for each value of $x$. }

\begin{figure}
	\includegraphics[scale=0.6]{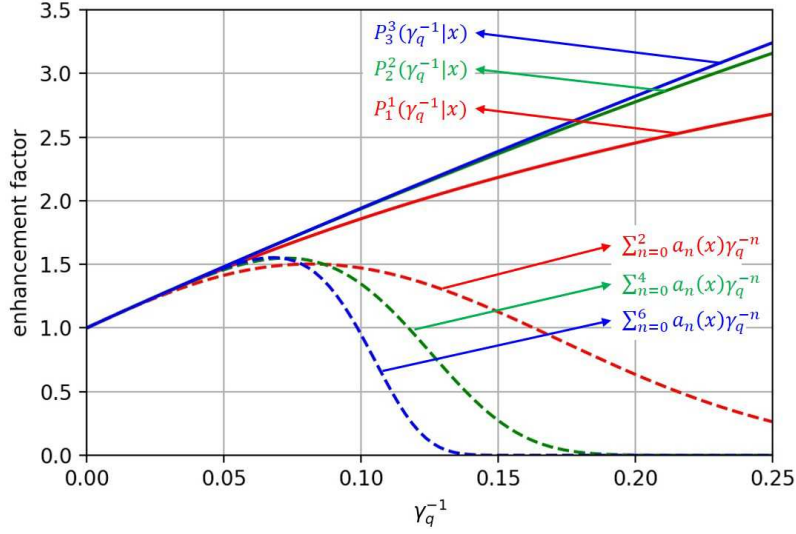}
	\caption{\label{fig:fig_AppC}{Enhancement factor $\exp(-S_{0}(\gamma_{q}^{-1}\protect\neq0))/\exp(-S_{0}(\gamma_{q}^{-1}=0))$
			as a function of $\gamma_{q}^{-1}$ for the case with $y=1$ and $x=2$.
			The solid lines were obtained with the Pad\'e approximants $P_{1}^{1}(\gamma_{q}^{-1}|x)$,
			$P_{2}^{2}(\gamma_{q}^{-1}|x)$, and $P_{3}^{3}(\gamma_{q}^{-1}|x)$;
			and the dashed lines with the polynominals $\sum_{n=0}^{2}a_{n}(x)\gamma_{q}^{-n}$,
			$\sum_{n=0}^{4}a_{n}(x)\gamma_{q}^{-n}$, and $\sum_{n=0}^{6}a_{n}(x)\gamma_{q}^{-n}$.
			According to the recurrence relations (\ref{eq:rec1}), (\ref{eq:rec2}),
			and (\ref{eq:rec3}), $P_{1}^{1}(\gamma_{q}^{-1}|x)$ is obtained
			from $\sum_{n=0}^{2}a_{n}(x)\gamma_{q}^{-n}$, $P_{2}^{2}(\gamma_{q}^{-1}|x)$
			from $\sum_{n=0}^{4}a_{n}(x)\gamma_{q}^{-n}$, and $P_{3}^{3}(\gamma_{q}^{-1}|x)$
			from $\sum_{n=0}^{6}a_{n}(x)\gamma_{q}^{-n}$, respectively.}}
\end{figure}

Pad\'e approximants are advantageous over polynomials,
as exemplified in Fig.~\ref{fig:fig_AppC} for the case with $y=1$
and $x=2$. In Fig.~\ref{fig:fig_AppC}, the enhancement factor $\exp(-S_{0}(\gamma_{q}^{-1}\neq0))/\exp(-S_{0}(\gamma_{q}^{-1}=0))$
is calculated with polynomials and diagonal PAs of various orders.
The polynomials and the PAs give nearly the same values until $\gamma_{q}^{-1}$
reaches 0.06, over which the polynomials quickly decrease to zero,
implying a quenching of DA-SPP with the assisting field. Such behavior
is unphysical and can be attributed to the polynomials' divergent
nature, which is usually more severe for higher orders. As a result,
the highest-order polynomial $\sum_{n=0}^{6}a_{n}(x)\gamma_{q}^{-n}$
produces more accurate values than the lowest-order polynomial $\sum_{n=0}^{2}a_{n}(x)\gamma_{q}^{-n}$
when $\gamma_{q}^{-1}<0.06$, but it is quicker to exhibit the anomalous
behavior as $\gamma_{q}^{-1}$ increases further. In contrast, the
PAs produce reasonable values, implying a monotonous enhancement of
DA-SPP with the assisting field. Furthermore, their convergence is
very good: the line obtained with $P_{2}^{2}(\gamma_{q}^{-1}|x)$
differs little from that with $P_{3}^{3}(\gamma_{q}^{-1}|x)$. Although
the PAs yield reasonable values, it is hardly possible to rigorously
estimate their accuracy if the corresponding polynomials diverge as
in our case \cite{Bender2013Advanced}. Not having a better technique
to handle such situation, we resort to the Pad\'e approximation method.

\bibliography{refs_SFQED}

\end{document}